\documentclass[a4paper,11pt]{article}
\usepackage{pos}
\newcommand{\be}{\begin{equation}}
\newcommand{\ee}{\end{equation}}
\newcommand{\ba}{\begin{eqnarray}}
\newcommand{\ea}{\end{eqnarray}}

\title{QCD-based estimate of direct $CP$ asymmetry \\in 
charm decays}
\ShortTitle{QCD-based estimates of direct $CP$ asymmetry...}

\author*[a]{ Alexander Khodjamirian}

\affiliation[a]{Center for Particle Physics Siegen  (CPPS),\\ Theoretische
  Physik 1, Universit\"at Siegen,\\D-57068 Siegen, Germany}

\emailAdd{khodjamirian@physik.uni-siegen.de}

\abstract{I  discuss the calculation  of the
direct CP-asymmetry in $D\to \pi^+\pi^-$ and $D\to K^+K^- $ decays 
with the method of QCD light-cone
sum rules. The main result is the upper limit for the difference of
the two asymmetries $\left|\Delta a_{CP}^{dir}\right|  < 0.020\pm
0.003\%$ which is significantly smaller than the recent measurement of
this quantity by the LHCb collaboration.}

\FullConference{%
  *** 10th International Workshop on Charm Physics (CHARM2020), ***\\
  *** 31 May - 4 June, 2021 ***\\
  *** Mexico City, Mexico - Online ***
}

\begin{document}
\maketitle

\section{Introduction}
The $CP$ asymmetry measured by LHCb collaboration
\cite{LHCb:2019hro} in the single Cabibbo suppressed decays
$D\to \pi^+\pi^-$ and $D\to K^+K^- $ remains a challenge
for the theory. A quantitative estimate of this asymmetry  in the
Standard Model (SM)
involves hadronic matrix elements with two energetic mesons in the final
state that are not accessible yet with the  lattice QCD methods. 
Hence, even approximate estimates of the $CP$ asymmetry obtained 
using  other QCD based methods  are useful, in order to trace or at least
constrain the possible beyond SM contributions (see e.g., \cite{Chala:2019fdb}).

Here I will discuss a calculation \cite{Khodjamirian:2017zdu} of the
direct CP-asymmetry in $D\to \pi^+\pi^-$ and $D\to K^+K^- $ decays 
with the method of light-cone
sum rules (LCSRs).  As shown in section 2,
for each of these decays it is sufficient to calculate the CKM suppressed part of the
decay amplitude with the penguin 
topology. In section 3, I briefly explain the idea of the LCSR method
in  a  simplified version with a two-point correlation
function. Section 4 presents the actual LCSR for the penguin hadronic
matrix element, obtained with a more involved procedure using  
the three-point correlation function. 
We transfer to  charm decays the method of LCSRs 
for $B\to \pi\pi$ decays initiated in \cite{Khodjamirian:2000mi} and 
developed further in \cite{Khodjamirian:2003eq,Khodjamirian:2005wn}. 
Our numerical
result  for the upper bound of  the difference of
direct CP asymmetries $\Delta a_{CP}$ is compared with the 
LHCb measurement \cite{LHCb:2019hro}. In conclusion, I discuss
the uncertainties and perspectives of our method.

\section{Relating direct CP-asymmetry to the penguin amplitude}

At the quark level, the single Cabibbo suppressed decays $D\to\pi\pi$ and $D\to K\bar{K}$ are generated  by the effective Hamiltonian
\ba
H_{eff}= \frac{G_{F}}{\sqrt{2}}\Big\{
\lambda_d\big(c_1~O_{1}^d+c_2~O_{2}^d\big) +
\lambda_s\big(c_1~O_{1}^s+c_2~O_{2}^s\big) 
-\lambda_b\!\!\!\!\sum\limits_{i=3,...,6,8g} c_{i} O_{i}
\Big\}
\,,
\label{eq:Heff}
\ea
where the operators
%
$ O_1^d =
\left(\bar{u}\Gamma_\mu d\right) 
\left(\bar{d}\Gamma^\mu c\right)
$\,,~
$
O_2^d =
\left(\bar{d}\Gamma_\mu d\right) 
\left(\bar{u}\Gamma^\mu c\right)
$
%
are multiplied with the Wilson coefficients $c_{1,2}$ and 
in $O^s_{1}\,, ~O^s_{2}$  the
$d$ quark is replaced by $s$ quark. Here,  $\lambda_{q}=V_{uq}V_{cq}^{*}, ~~   (q=d,s,b)$ are the relevant 
combinations of the CKM parameters.

Separating the contributions of the ${\cal O}^d_{1,2}$ and 
${\cal O}^s_{1,2}$ operators and introducing a compact notation
$${\cal O}^q\equiv \sum_{i=1,2}c_iO_i^q\, ~~(q=d,s), $$ 
we have for the decay amplitudes 
\footnote{ In what follows, we neglect the
contributions of the  effective operators $O_{i\geq 3}$ since
$c_{i\geq 3}\ll c_{1,2}$.}
(in the units of $G_F/\sqrt{2}$):
\begin{eqnarray}
&&A(D^0\to \pi^+\pi^-)=
 \lambda_d\langle \pi^+\pi^-|{\cal O}^{d}|D^0\rangle +
\lambda_s\langle \pi^+\pi^-|{\cal O}^{s}|D^0\rangle \,, 
\label{eq:ADpipi}\\
&&A(D^0\to K^+K^-)=
\lambda_s\langle K^+K^-|{\cal O}^{s}|D^0\rangle +
 \lambda_d \langle K^+K^-|{\cal O}^{d}|D^0\rangle \,. 
\label{eq:ADKK}
\end{eqnarray}
Note that both decay amplitudes contain hadronic matrix elements  with "penguin topology'' 
\begin{equation}
\langle \pi^+\pi^-|{\cal O}^{s}|D^0\rangle\equiv {\cal P}_{\pi\pi}^s \,,~~~~
\langle K^+K^-|{\cal  O}^{d}|D^0\rangle\equiv{\cal P}_{KK}^d\,,
\label{eq:peng}
\end{equation}
in which the quark 
operator contains   a $\bar{q}q$ pair with  a flavour $q=s$ or $d$
absent in the valence content of the initial and final hadronic states.
This definition is somewhat more general than specifying  certain quark-flow 
diagrams (quark topologies).

Furthermore, using the CKM unitarity in SM: 
$\sum_{q=d,s,b} \lambda_q=0\,,$ 
we find it convenient to eliminate 
$\lambda_d=-(\lambda_s+ \lambda_b)\,,$
so that the amplitudes (\ref{eq:ADpipi}) and (\ref{eq:ADKK}) are written in the form
\begin{eqnarray}
A(D^0\to \pi^+\pi^-)&=& 
- \lambda_s {\cal A }_{\pi\pi}
\Big\{1 +\frac{\lambda_b}{\lambda_s}\Big(1+  r_{\pi}\exp(i\delta_{\pi})\Big)\Big\}\,, 
\label{eq:Dpipi2} 
\\ 
A(D^0\to K^+ K^-)&=& 
 \lambda_s{\cal A }_{KK}
\Big\{ 1-\frac{\lambda_b }{\lambda_s} r_{K}\exp(i\delta_{K})\Big\}\,, 
\label{eq:DKK2} 
\end{eqnarray}
with the notation
\begin{eqnarray}
{\cal A }_{\pi\pi}=\langle \pi^+\pi^-|{\cal O}^{d}|D^0\rangle-
\langle \pi^+\pi^-|{\cal O}^{s}|D^0\rangle\,,~~
{\cal A }_{KK}=\langle K^+K^-|{\cal O}^{s}|D^0\rangle-
\langle K^+K^-|{\cal O}^{d}|D^0\rangle\,,
\label{eq:Acal}
\end{eqnarray}
and 
\begin{eqnarray}
 r_{\pi}=\left|\frac{{\cal P}_{\pi\pi}^s}{{\cal A}_{\pi\pi}}
\right|\,, ~~
 r_{K}=\left|\frac{{\cal P}_{KK}^d}{{\cal A}_{KK}}\right|\,, ~
\delta_{\pi (K)}= arg[{\cal P}_{\pi\pi(KK)}^{s(d)}]-arg[{\cal A}_{\pi\pi(KK)}]\,.
\label{eq:ratios}
\end{eqnarray}
In (\ref{eq:Dpipi2}) and (\ref{eq:DKK2}) only the 
contributions proportional to $\lambda_b$, with
$\mbox{Im}(\lambda_b)\neq 0$,  contain the CP-violating phase.
The conditions for a nonvanishing direct $CP$ asymmetry are clearly fulfilled:
both decay amplitudes consist  of  two  parts
with different  weak  and strong  phases. 
 Defining this  asymmetry as 
\begin{eqnarray}
&&a_{CP}^{dir}(f)=\frac{\Gamma(D^0\to f)-\Gamma(\bar{D}^0\to \bar{f})}{
\Gamma(D^0\to f)+\Gamma(\bar{D}^0\to \bar{f})}\,, ~~~f=\pi^+\pi^-,\,K^+K^-\,,
\label{eq:aCP} 
\end{eqnarray}
and using (\ref{eq:Dpipi2}) and (\ref{eq:DKK2}), we obtain:
\begin{eqnarray}
&&a_{CP}^{dir}(K^+K^-)=\frac{-2 r_{b} r_{K}\sin \delta_{K} \sin \gamma}{1 - 
2 r_{b} r_{K}\cos\gamma \cos\delta_{K} +
 r_{b}^2r_{K}^2}\,,
\label{eq:AcpKK} \\
&&a_{CP}^{dir}(\pi^+\pi^-)=\frac{2 r_{b} r_{\pi}\sin \delta_{\pi} \sin \gamma}{1
+2 r_{b}\cos\gamma(1+ r_{\pi}\cos\delta_\pi)+
 r_{b}^2(1+2 r_{\pi}\cos\delta_{\pi}+ r_{\pi}^2)}\,,
\label{eq:Acp2pi}
\end{eqnarray}
were the ratio of the CKM matrix elements is parameterized as  
\begin{equation}
 \frac{\lambda_b}{\lambda_s}\equiv  r_be^{-i\gamma},~~
  r_b=\Bigg|\frac{V_{ub}V_{cb}^*}{V_{us}V_{cs}^*}\Bigg|\,.
\end{equation}
Furthermore, it is important that 
$|\lambda_b|\ll |\lambda_{s,d}|$, hence both equations 
(\ref{eq:AcpKK}) and (\ref{eq:Acp2pi}) can be 
expanded in $r_b$ retaining the first power.
As well known, a more "clean" observable (after the time integration)
than the individual $CP$ asymmetries (\ref{eq:AcpKK}) and (\ref{eq:Acp2pi})
is their difference: 
\begin{eqnarray}
\Delta a_{CP}^{dir}&=& a_{CP}^{dir}(K^+K^-)-a_{CP}^{dir}(\pi^+\pi^-)
\nonumber \\
&=& 
-2 r_b \sin\gamma( r_K \sin\delta_K+ r_\pi \sin \delta_\pi) +O( r_b^2)\,,
\label{eq:deltaa}
\end{eqnarray}
where the hadronic input is reduced to the two ratios and two phase 
differences defined in  (\ref{eq:ratios}).
At first sight, it seems that  we have 
unnecessarily complicated our task  defining the combinations ${\cal
  A}_{\pi\pi}$ and $ {\cal A}_{KK}$ of the hadronic matrix elements.
In fact,  the key point is that, due to
the smallness of $|\lambda_b|$, from (\ref{eq:Dpipi2}) and (\ref{eq:DKK2}) we have, to a good approximation,
$$- \lambda_s{\cal A}_{\pi\pi}\simeq A(D^0\to \pi^+\pi^-)\,,~~~
\lambda_s{\cal A}_{KK}\simeq A(D^0\to K^+K^-)\,.$$ 
Thus, the 
denominators of the ratios $r_K$ and $r_\pi$ are obtained from
the experimentally measured widths of these two decays. 
The estimate  of direct $CP$ asymmetries (\ref{eq:AcpKK}) 
and (\ref{eq:Acp2pi}) in SM is then reduced to the calculation of
the two hadronic matrix elements ${\cal P}_{\pi\pi}^{s}$ and  
${\cal P}_{KK}^{d}$ defined in (\ref{eq:peng}).

\section{ Outline of the LCSR method}
\begin{figure}
\begin{center}
\includegraphics[scale=1.0]{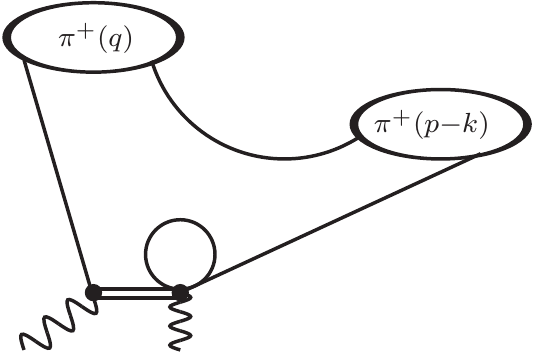}\hspace{1cm}
\includegraphics[scale=1.0]{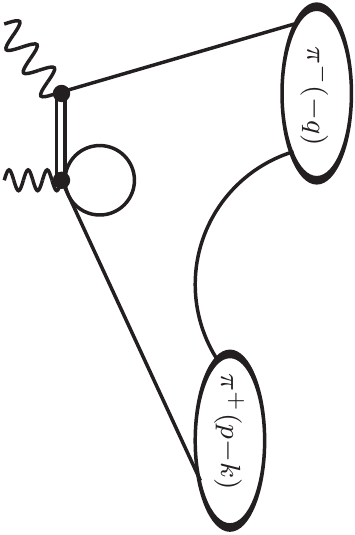}
\caption{The diagram illustrating OPE for the 
correlation function (\ref{eq:corr2pt}) for the spacelike (left) and
timelike (right) invariant variable $P^2=(p-k-q)^2$. Shown are the $s$-quark
loop and the virtual $c$ quark (double line). The wavy lines denote external momenta
$(p-q)$ and $k$ flowing, respectively, to the interpolating current and from the weak
operator vertex. }
\label{fig1}
\end{center}
\end{figure}
These  hadronic matrix elements 
were calculated in \cite{Khodjamirian:2017zdu}, employing the LCSR
method formulated and applied to the $B\to \pi\pi$ decay in 
\cite{Khodjamirian:2000mi,Khodjamirian:2003eq,Khodjamirian:2005wn}. 
Let me first outline  a somewhat simplified version
of that method, in which the starting object is
a two-point correlation function.  Considering e.g., the
$D\to \pi\pi$ decay, we introduce:
\begin{equation}
\Pi_{\pi\pi}( (p-q)^2,P^2)=i \int \!d^4x \,e^{-i(p-q)x}\! 
\langle \pi^+(p-k) |T\{ {\cal O}^{s}(0)
j_{5}^{(D)}(x)\}|\pi^+(q)\rangle\,,
\label{eq:corr2pt}
\end{equation}
where $ j_{5}^{(D)}=m_c\bar{c}i\gamma_5 u$ is the $D^0$-meson
interpolating current~\footnote{ The idea to describe a  nonleptonic decay amplitude of heavy meson 
using a two-point correlator and OPE goes back to \cite{Blok:1992na}}.
 Its product with the four-quark effective operator 
is sandwiched between the initial and final on-shell pion states, so 
that $(p-k)^2=q^2=m_\pi^2$ .  Note that, as explained in detail in \cite{Khodjamirian:2000mi},  an artificial
four-momentum $k$ is added, flowing from the weak operator vertex. The 
invariant variables $(p-q)^2$ and $P^2=(p-k-q)^2$  
are spacelike and large, 
\begin{equation}
|P^2|,\, |(p-q)^2|\gg \Lambda_{QCD}^2
\label{eq:spcl}
\end{equation}
and, for simplicity, $p^2=0$, $k^2=0$ are chosen. In this region of
invariant variables 
the propagating $c$~quark is far off shell and 
the operator product expansion (OPE) near the light-cone $x^2\sim 0$
is valid for (\ref{eq:corr2pt}), starting from the bilocal
light-quark-antiquark operator, schematically
\begin{equation} 
T\{ {\cal O}^{s}(0) j_{5}^{(D)}(x)\}=C_\Gamma(x^2)\bar{u}(x)\Gamma u(0)+\dots\,,
\label{eq:ope}
\end{equation}
where the ellipsis denotes all other operators and $\Gamma$ denotes the
relevant Dirac structure. Using
(\ref{eq:ope}), we reduce (\ref{eq:corr2pt}) to  a convolution 
\begin{eqnarray}
\Pi_{\pi\pi}( (p-q)^2,P^2)&=&\Pi^{(OPE)}_{\pi\pi}( (p-q)^2,P^2)
\nonumber\\
&=&
i \int \!d^4x \,e^{-i(p-q)x} C_\Gamma(x^2)
\langle \pi^+(p-k) |T\{\bar{u}(x)\Gamma u(0)\}|\pi^+(q)\rangle +\dots\,.
\label{eq:corr2ptOPE}
\end{eqnarray}
Diagrammatically, it is shown  in 
Fig.~1, where 
the short-distance part contains the $s$-quark loop and $c$-quark propagator.
The  long-distance part emerges as a  pion-to-pion matrix element 
of a bilocal quark-antiquark operator. In the local limit , $x\to 0$,
this matrix element reduces to  a certain pion form factor at
spacelike momentum, $P^2<0$.

Let us assume  that we are able to obtain
$\Pi_{\pi\pi}( (p-q)^2,P^2)$ from the OPE  (\ref{eq:corr2ptOPE}), computing
the short distance part perturbatively and combining it with a known
parameterization  of the hadronic matrix element. The latter is a process
independent quantity, hence, it is conceivable that it can be
determined independently, e.g. inferred from a 
dedicated LCSR    \footnote{ Note that by its
  structure the long distance part in (\ref{eq:corr2ptOPE}) resembles 
matrix elements determining the 
generalized parton distributions of the pion, see e.g. the review \cite{Diehl:2003ny}.} .
Then, the next  step is an analytical continuation 
of the correlation function (\ref{eq:corr2pt}) from large spacelike
to large timelike values of $P^2$, more specifically to $P^2=m_D^2$, 
so that  the invariant mass of the dipion state is equal to the $D$-meson mass,
while the spacelike  variable $(p-q)^2$ remains fixed. We encounter 
the vacuum-to-dipion matrix element 
\begin{equation}
\Pi_{\pi\pi}( (p-q)^2,m_D^2)=i \int \!d^4x \,e^{-i(p-q)x}\! 
\langle \pi^+(p-k) \pi^-(-q)|T\{ {\cal O}^{s}(0) j_{5}^{(D)}(x)\}|0\rangle\,.
\label{eq:corr2ptt}
\end{equation}
The timelike and spacelike asymptotics of the same correlation function
coincide, hence from (\ref{eq:corr2ptOPE})
 we have:
\begin{equation}
\Pi_{\pi\pi}( (p-q)^2,m_D^2)\simeq \Pi^{(OPE)}_{\pi\pi}(
(p-q)^2,P^2=m_D^2)\,.
\label{eq:Pidual}
\end{equation}

Note that a complex phase can be generated by the continuation of the 
OPE result, in our case, e.g., due to the discontinuity of $s$-quark
loop in the timelike region.  That this phase
reproduces  the strong phase of the final state dipion interaction is a rather strong
assumption. A weaker version of (\ref{eq:Pidual})  is the equality of the
absolute values of both sides. Another form of the relation between timelike and spacelike
asymptotics  emerges if we write down dispersion relations in the variable $P^2$ for both
sides of (\ref{eq:corr2ptOPE}). Equating them at large $P^2\to
-\infty$ yields the asymptotic equality  of the hadronic and OPE
spectral densities: $\mbox{Im} \Pi_{\pi\pi}( (p-q)^2,s)\simeq \mbox{Im}\Pi^{(OPE)}_{\pi\pi}(
(p-q)^2,s)$ at $s\to \infty$, which is a manifestation of  the local quark-hadron duality.
To demonstrate that the approximation (\ref{eq:Pidual}) is valid for simpler hadronic
matrix elements such as the pion electromagnetic form factor, we refer to 
\cite{Cheng:2020vwr},  where  the absolute values of the pion timelike
and  spacelike form factors (the former measured and the
latter calculated from a dedicated LCSR) are compared. We observe that
local duality works at large  dipion invariant masses, starting from
2.5 - 3.0 GeV, that is, in the ballpark of the $D$ meson mass.

The last step is to employ the dispersion relation for the amplitude  
$\Pi_{\pi\pi}( (p-q)^2,m_D^2)$ in the variable $(p-q)^2$. Inserting the
total set of hadronic states with $D$-meson quantum numbers, we obtain:
\begin{equation}
\Pi_{\pi\pi}( (p-q)^2,m_D^2)=\frac{f_Dm_D^2
\langle \pi^+(p) \pi^-(-q)|{\cal O}^{s}
|D(p-q)\rangle}{m_D^2-(p-q)^2} +\int\limits_{s_h}^{\infty}\, ds\frac{\rho^{(D)}_h(s)}{s-(p-q)^2}\,,
\label{eq:corr2ptdisp}
\end{equation}
where  the ground-state contribution contains the $D\to \pi\pi$ 
matrix element of  ${\cal O}^{s}$ 
multiplied by the $D$-meson decay constant and
$\rho^{(D)}_h(s)$  denotes the hadronic spectral density of excited
and  continuum states. Importantly,  due
to the choice $P^2=(p-q-k)^2=(p-q)^2=m_D^2$, 
the fictitious momentum $k$ vanishes in the pole term . 
Matching this dispersion relation to  the OPE and applying the
(semi-local) quark-hadron duality in the $D$-meson channel:
\begin{equation}
\int\limits_{s_h}^{\infty}\!\!ds\frac{\rho^{(D)}_h(s)}{s-(p-q)^2}=
\frac1\pi\int\limits_{s_0}^{\infty}\!\! ds\,\frac{
  \mbox{Im}\Pi^{(OPE)}_{\pi\pi}( s,m_D^2)}{s-(p-q)^2}\,,
 \label{eq:lcsrdual}
\end{equation}
we obtain, after the standard  Borel transformation $(p-q)^2\to M^2$, 
the LCSR for the $D\to \pi\pi$ amplitude with penguin topology:
\begin{equation}
{\cal P}_{\pi\pi}^s=\frac{e^{m_D^2/M^2}}{ f_Dm_D^2}\frac1\pi\int
\limits_{m_b^2}^{s_0^D}\!\! ds\, e^{-s/M^2} \mbox{Im}\Pi^{(OPE)}_{\pi\pi}( s,m_D^2)\,.
\label{eq:lcsr}
\end{equation}
Replacing in the correlation function (\ref{eq:corr2ptOPE}) $\pi\to K$
and ${\cal O}^s\to {\cal O}^d$, we repeat the same procedure and
obtain the LCSR for the penguin amplitude ${\cal P}_{KK}^d$.

The procedure based on the two-point
correlation function was presented here mainly to illustrate two main
elements of the LCSR method for weak nonleptonic decays: the transition
to timelike region and the sum rule in the $D$ meson channel.  
The actual calculation of $D\to \pi\pi$ amplitudes with  penguin
topology in \cite{Khodjamirian:2017zdu} essentially 
uses both these elements, however, starts from
 a  three-point  correlation function,  
following  \cite{Khodjamirian:2000mi}, and,  more specifically , using
 the calculation of $B\to 2\pi$ amplitudes 
with the $c$-quark penguin topology \cite{Khodjamirian:2003eq}.
The additional operator in the 
correlation function  is the pion interpolating current. It is needed because the pion-to-pion matrix element 
entering OPE such as the one in (\ref{eq:corr2ptOPE}) 
is not directly accessible. One calculates this matrix element
using an additional QCD sum rule in the pion channel.

Before we turn to this calculation in more details,
let me parenthetically mention 
that the two-point correlation function was employed 
in \cite{Khodjamirian:2005wn} 
to obtain LCSR for the annihilation contribution with hard-gluon exchange 
in $B\to\pi\pi$ decays. In this case  the pion-to-pion
matrix element was factorized into  two 
light-cone distribution amplitudes of the pion convoluted with
perturbatively calculated short-distance part.

\section{Penguin amplitudes from LCSR}

As explained in the previous section, to access the penguin amplitude 
${\cal P}_{\pi\pi}^s$ in  $D\to \pi^+\pi^-$ decay, we introduce the 
three-point  vacuum-to-pion correlation function 
%
\begin{eqnarray}
F_\alpha(p,q,k)=i^{\,2} \int \!d^4x \,e^{-i(p-q)x}\! \int \! d^4y \, e^{-i(p-k)y}
\langle 0 |T\{ j_{\alpha 5}^{(\pi)}(y){\cal O}^{s}(0)
j_{5}^{(D)}(x)\}|\pi^+(q)\rangle
\nonumber\\=F((p-k)^2,(p-q)^2,P^2)(p-k)_\alpha+ \dots\,,
\label{eq:corr3pt}
\end{eqnarray}
where  $j_{\alpha 5}^{(\pi)}=\bar{u}\gamma_\alpha\gamma_5d$ is the
pion interpolating current and we isolate the relevant invariant
amplitude.  The light-cone OPE of this function and of the analogous 
one for $D\to K^+K^-$ (obtained replacing $d\to s, s\to d$ and $\pi \to K$ in
the above) 
are accessible in  terms of the pion and, respectively, kaon DAs 
of the growing twist, similar to 
the simpler correlation functions used to determine the $D\to \pi$  and
$D\to  K$ form
factors
(see e.g. \cite{Khodjamirian:2009ys} ). This expansion is valid 
in the region (\ref{eq:spcl}),  assuming
that the new variable $(p-k)^2$ is also spacelike and large.
Transforming the four-quark operators with a colour Fierz transformation:
\begin{eqnarray}
c_1 O_1^s+c_2O^s_2= 2c_1\widetilde{O}_2^s+ 
\left(\frac  {c_1}{3}+c_2\right)O^s_2\,,~~
\widetilde{O}_2^s= \left(\bar{s}\Gamma_\mu \frac{\lambda^a}{2}s\right) 
\left(\bar{u}\Gamma^\mu\frac{\lambda^a}{2} c\right)\,,
\label{eq:operators}
\end{eqnarray}
we realize that the colour-octet operator provides the dominant
contribution. Hence, up to NLO corrections,
${\cal P}_{\pi\pi}^s=2c_1\langle \pi^+\pi^-|
\widetilde{O}_2^s|D^0\rangle$, and, 
analogously, ${\cal P}^d_{KK}= 2c_1\langle K^+K^-|\widetilde{O}_2^d|D^0\rangle$.
The relevant OPE diagrams are presented and discussed in detail 
in \cite{Khodjamirian:2017zdu} (see also \cite{Khodjamirian:2003eq}),
They are reduced to the short-distance parts (loop and propagators)
convoluted with the pion DAs of growing twist and  multiplicity. In
particular, the $s$-quark pair with a small virtuality, emitted from the
weak vertex  is not described by the loop diagram, but forms a part of 
four-particle pion DAs. Such contributions are suppressed by
inverse powers of large scales and are neglected.

 The OPE result for the correlation function (\ref{eq:corr3pt}) 
is equated to  the dispersion relation in
the variable $(p-k)^2$. Furthermore,  using the  quark-hadron 
(semilocal) duality we isolate the contribution
of the pion.  After  Borel transformation, $(p-k)^2\to M^{'2}$, we
obtain
\begin{equation}
\Pi_{\pi\pi}((p-q)^2,P^2)=\left(\frac{-i}{ f_\pi}\right)\frac1\pi\int
\limits_{0}^{s_0^\pi}\!\! ds\, e^{-s/M^{'2}} \mbox{Im}_s
F^{(OPE)}_{\pi\pi}(s, (p-q)^2,P^2 )\,,
\label{eq:lcsr2}
\end{equation}
where, as usual in the sum rules for a heavy-meson to pion transitions,
the chiral symmetry  is adopted with a massless pion.
The above sum rule yields the pion-to-pion correlation function
defined in (\ref{eq:corr2pt}) in the spacelike region  (\ref{eq:spcl}). 
Accordingly, the subsequent steps for this correlation function 
repeat the ones described
in the previous section. The resulting LCSR for 
${\cal P}^s_{\pi\pi} $ has the same form as 
(\ref{eq:lcsr}) but with a double integral and  double imaginary part in the variables
$(p-k)^2$, $(p-q)^2$. 
Explicit expressions of this sum rule and its analog for 
${\cal P}^d_{KK}$ are given in  \cite{Khodjamirian:2017zdu}.

 \section{Results and discussion}

The numerical analysis of  LCSRs for the hadronic matrix elements 
${\cal P}_{\pi\pi}^s$ and ${\cal P}_{KK}^d  $ needs inputs of
three types: (1) the QCD parameters such as $\alpha_s$, the quark masses
$m_c$ and 
$m_s$, (hence, we can assess the $SU(3)_{fl}$-symmetry violation), whereas  $m_u=m_d=0$;
(2) the set of pion and kaon DAs of twist 2,3 and, finally (3) the
Borel parameter intervals and effective thresholds in channels of the pion 
($M'$ and $s_0^\pi$, $s_0^K$) and $D$-meson ($M$ and $s_0^D$). The adopted
values of all these parameters, including also  the effective
coefficient  $c_1$,  can be found in \cite{Khodjamirian:2017zdu}.
Our final numerical results obtained from LCSR are
\be
\frac{G_F}{\sqrt{2} }|{\cal P}_{\pi\pi}^{s}|= (1.96 \pm 0.23) \times 10^{-7} \mbox {GeV}\,,
~~~
\frac{G_F}{\sqrt{2} }|{\cal P}_{KK}^{d}|= (2.86 \pm 0.56) \times 10^{-7}\mbox {GeV}\,,
\label{eq:res}
\ee
The quoted uncertainties are only parametrical. 
Using  experimentally measured branching fractions \cite{PDG}
of $D^0\to \pi^+\pi^-$ and $D^0\to K^+K^-$, we obtain
\begin{eqnarray}
r_\pi = \frac{|{\cal P}_{\pi\pi}^{s}|}{|{\cal A}_{\pi\pi}|}= 0.093 \pm 0.011\,, \qquad
r_K = \frac{|{\cal P}_{KK}^{d}|}{|{\cal A}_{KK}|}= 0.075 \pm 0.015\,.
\nonumber
\end{eqnarray}
The direct CP asymmetries are obtained using
the CKM parameter averages from \cite{PDG}: 
$ r_b\,sin\gamma= 0.64\times 10^{-3} $.
The resulting upper limits on the direct asymmetries and their
difference (independent of strong phases) are
\begin{eqnarray} \label{eq:AcpNum1}
&&\left |a_{CP}^{dir}(\pi^-\pi^+)\right | < 0.012\pm 0.001\%,~~ 
\left |a_{CP}^{dir}(K^-K^+) \right| < 0.009\pm 0.002 \%,
\nonumber \\
&&\left|\Delta a_{CP}^{dir}\right|  < 0.020\pm 0.003\%\,.
\label{eq:acpres}
\end{eqnarray}
The latter  turns out to be substantially smaller than the 
most recent 
measurement by LHCb collaboration  \cite{LHCb:2019hro}:
\begin{equation} 
\Delta a^{dir}_{CP}=(-0.154 \pm 0.029)\,.
\label{eq:exp}
\end{equation}
Leaving aside this tension and its interpretation
(see e.g. \cite{Chala:2019fdb}),
let us discuss the accuracy of our prediction and the perspectives
of improving it.

The  accuracy of LCSR, (apart from the input parameter
variation within the adopted intervals) is determined by the missing
higher-twist terms (starting from twist 4). In future, it is possible 
to add them to OPE, but they are usually small,
as e.g., in the LCSRs for $D\to \pi,K$ form factors.
The  $O(\alpha_s^2)$ corrections not included in our calculation
are probably also small, but technically difficult to compute.
 Furthermore, since
we  used  analytical expressions from \cite{Khodjamirian:2003eq},
certain  terms of $O(s^{\pi,K}_0/m_D^2)$ are neglected, restoring them
demands dedicated calculation. It is however not conceivable that
 adding higher twists, 
NLO terms and neglected power corrections to the LCSR will
shift the result in (\ref{eq:acpres})  by a large factor.

The potentially most important  source  of uncertainty not fully accounted
in (\ref{eq:acpres})
is the use of  local quark-hadron duality. The
timelike scale $m_D^2$ might still be somewhat  small for  the onset of  asymptotics,
and an enhancement due to intermediate scalar resonances $f_0(J^P=0^+)$ decaying to
$\pi\pi$ and $K\bar{K}$ is not excluded (see
e.g., \cite{Soni:2019xko}, \cite{Schacht:2021jaz}).
One possibility to study the effect of resonances is to
match  the LCSR calculation at spacelike $P^2$ to the dispersion relation saturated by
resonances. This will introduce a certain  model dependence
\footnote{ A similar
approach was used in the study of nonlocal effects in the exclusive 
$b\to s\ell\ell$ decays \cite{Khodjamirian:2010vf}.}.
Another perspective is to extend the applications of the LCSR method to  other
hadronic decays of bottom and charmed hadrons, e.g.
we plan to use it for the two-body  decays of  heavy baryons \cite{new}. 

{\bf Acknowledgements}

 I thank Hua-Yu Jiang for a useful discussion.
This research was supported by the DFG (German Research Foundation) 
 under the grant 396021762 - TRR 257.


\begin{thebibliography}{99}


\bibitem{LHCb:2019hro}
R.~Aaij \textit{et al.} [LHCb],
Phys. Rev. Lett. \textbf{122} (2019) no.21, 211803
[arXiv:1903.08726 [hep-ex]].

\bibitem{Chala:2019fdb}
M.~Chala, A.~Lenz, A.~V.~Rusov and J.~Scholtz,
JHEP \textbf{07} (2019), 161
[arXiv:1903.10490 [hep-ph]].


\bibitem{Khodjamirian:2017zdu}
A.~Khodjamirian and A.~A.~Petrov,
Phys. Lett. B \textbf{774} (2017), 235-242
[arXiv:1706.07780 [hep-ph]].



\bibitem{Khodjamirian:2000mi}
A.~Khodjamirian,
Nucl. Phys. B \textbf{605} (2001), 558-578
[arXiv:hep-ph/0012271 [hep-ph]].

\bibitem{Khodjamirian:2003eq}
A.~Khodjamirian, T.~Mannel and B.~Melic,
Phys. Lett. B \textbf{571} (2003), 75-84
[arXiv:hep-ph/0304179 [hep-ph]].

\bibitem{Khodjamirian:2005wn}
A.~Khodjamirian, T.~Mannel, M.~Melcher and B.~Melic,
Phys. Rev. D \textbf{72} (2005), 094012
[arXiv:hep-ph/0509049 [hep-ph]].

\bibitem{Blok:1992na}
B.~Blok and M.~A.~Shifman,
Nucl. Phys. B \textbf{389} (1993), 534-548
[arXiv:hep-ph/9205221 [hep-ph]].


\bibitem{Diehl:2003ny}
M.~Diehl,
Phys. Rept. \textbf{388} (2003), 41-277
[arXiv:hep-ph/0307382 [hep-ph]].

\bibitem{Cheng:2020vwr}
S.~Cheng, A.~Khodjamirian and A.~V.~Rusov,
Phys. Rev. D \textbf{102} (2020) no.7, 074022
[arXiv:2007.05550 [hep-ph]].

\bibitem{Khodjamirian:2009ys}
A.~Khodjamirian, C.~Klein, T.~Mannel and N.~Offen,
Phys. Rev. D \textbf{80} (2009), 114005
[arXiv:0907.2842 [hep-ph]].

\bibitem{PDG}
P.~A.~Zyla \textit{et al.} [Particle Data Group], 
PTEP \textbf{2020} (2020) no.8, 083C01 



\bibitem{Soni:2019xko}
A.~Soni,
[arXiv:1905.00907 [hep-ph]].

\bibitem{Schacht:2021jaz}
S.~Schacht and A.~Soni,
[arXiv:2110.07619 [hep-ph]].


\bibitem{Khodjamirian:2010vf}
A.~Khodjamirian, T.~Mannel, A.~A.~Pivovarov and Y.~M.~Wang,
JHEP \textbf{09} (2010), 089
[arXiv:1006.4945 [hep-ph]].


\bibitem{new}
S.~Cheng, H.~Y.~ Jiang, A.~Khodjamirian and F.~S.~Yu, {\it work in progress}

\end{thebibliography}
\end{document}